\documentclass[preprint,12pt]{elsarticle}

\usepackage{amssymb}
 \usepackage{amsthm}
\usepackage{amsmath}
\usepackage{bbm}
\usepackage{comment}

\journal{ Physics Letters B}

\begin{document}

\begin{frontmatter}



\title{Branes and wrapping rules}


\author[ea]{Eric A. Bergshoeff}

\ead{E.A.Bergshoeff@rug.nl}

\author[fr]{Fabio Riccioni}

\ead{fabio.riccioni@roma1.infn.it}

\address[ea]{Centre for Theoretical Physics, University of Groningen, \\ Nijenborgh 4, 9747 AG Groningen, The
Netherlands}

\address[fr]{INFN Sezione di Roma, \\  Dipartimento di Fisica, Universit\`a di Roma ``La Sapienza'' \\ Piazzale Aldo Moro 2, 00185 Roma, Italy}

\begin{abstract}
We show that the branes of ten-dimensional IIA/IIB string theory
must satisfy, upon toroidal compactification, specific wrapping
rules in order to reproduce the number of supersymmetric branes that
follows from a supergravity analysis. The realization of these
wrapping rules suggests that IIA/IIB string theory contains
a whole
class of generalized Kaluza-Klein monopoles.

\end{abstract}


\end{frontmatter}


\section{Introduction}
\label{introduction}

It is by now well-understood  that branes form a crucial ingredient
of string theory. For instance, they have been used to calculate the
entropy of certain black holes \cite{Strominger:1996sh} and they are
at the heart of the AdS/CFT correspondence \cite{Maldacena:1997re}.
Often, the presence of a $p$-brane in string theory can be deduced
from the presence of a rank $p+1$-form potential in the
corresponding supergravity theory. It is a relatively new insight
that the potentials of a given supergravity theory are not only the
ones that describe the physical degrees of freedom of the
supermultiplet. It turns out that the supersymmetry algebra allows
additional high-rank potentials that do not describe any degree of
freedom but, nevertheless, play an important role in describing the
coupling of branes to background fields. For maximal supergravity theories,
the allowed U-duality
representations of these ``un-physical'' potentials have been
classified in \cite{Riccioni:2007au,Bergshoeff:2007qi,deWit:2008ta}.

A distinguishing feature of the un-physical potentials  is that,
when considered in different dimensions, they are not related to
each other by toroidal compactification. This is unlike the
``physical'' potentials, including the dual potentials, whose
numbers are fixed by the representation theory of the supersymmetry
algebra.  Indeed, all physical potentials are related by toroidal
compactification. Supergravity is therefore not complete in the
sense that the lower-dimensional supergravity theories, including
the un-physical potentials, do not follow from the reduction of the
ten-dimensional supergravity theory. It is this incomplete nature of
supergravity that will lead us to suggest
a class of
generalized Kaluza-Klein (KK) monopoles in string theory.

In this letter we will consider the supersymmetric branes of IIA/IIB
string theory compactified on a torus, which couple to the fields of
the corresponding maximal supergravities. As mentioned above these
fields do not only include the physical potentials, i.e.~the
$p$-forms with $0 \le p \le D-2$ but also the un-physical
potentials, i.e.~$D-1$-forms (which are dual to constant parameters)
and $D$-forms (that have no field strength). In
\cite{Bergshoeff:2011mh} we distinguished between {\it standard}
branes, i.e.~branes of co-dimension higher than 2, and {\it
non-standard} branes, i.e.~branes of co-dimension 2,1 and 0. While
standard branes are automatically classified because their number
coincides with the dimension of the U-duality representation of the
corresponding field, this is in general not true for the
non-standard branes. A prototype example are the 7-branes of IIB
string theory: although the supersymmetry algebra closes on an
$\text{SL}(2, \mathbb{R})$ triplet of 8-forms\,\footnote{Actually,
the situation in this case is slightly more subtle since the triplet
of 9-form curvatures of these potentials satisfies a non-linear
constraint. This is a general property of branes of co-dimension 2
which does not play a role in the present discussion.}, only two of
them are actually associated to supersymmetric branes
\cite{Bergshoeff:2006gs}: the D7-brane and its S-dual. At present,
it has not been worked out what the number
of supersymmetric non-standard branes is in a given dimension.

Recently, a step forward in this direction was performed in
\cite{Bergshoeff:2010xc,Bergshoeff:2011zk}. The strategy of these
papers was to analyse the structure of the gauge-invariant
Wess-Zumino (WZ) terms and to introduce the following brane criterion: a potential can be associated to a supersymmetric brane
if the corresponding gauge-invariant WZ term requires the
introduction of world-volume fields that fit within the bosonic
sector of a suitable supermultiplet. Decomposing in each dimension $D=10-d$
the U-duality representations in terms of T-duality representations as
  \begin{equation}\label{decomposition}
  {\rm U-duality}\ \supset\ {\rm SO}(d,d)\times \mathbb{R}^+
  \end{equation}
one can deduce how the tension $T$ of each brane scales with the string coupling constant $g_S$
in
terms of a number $\alpha$
\begin{equation}
T\ \sim \ (g_S)^\alpha\,.
\end{equation}
The value of  $\alpha$ follows from the $\mathbb{R}^+$-weight of the
corresponding potential. The analysis of the fields as T-duality
representations for each value of $\alpha$ reveals a remarkable
recurrence \cite{Bergshoeff:2010xc,Bergshoeff:2011zk} at least for
the highest values of $\alpha$. The fundamental fields, that is the
fields with $\alpha =0$, are in all cases a 1-form and a 2-form,
which transform respectively as a vector and a singlet under
T-duality.\footnote{In this letter we are only interested in gauge
fields, and we therefore do not consider scalars (which would couple
to instantons).} The RR fields, which have $\alpha =-1$, are in all
dimensions T-duality spinors of alternating chirality. Finally, the
solitonic fields, with $\alpha=-2$, belong to T-duality
representations corresponding to antisymmetric tensors of rank zero
to four (see \cite{Bergshoeff:2011zk} for the details).

While the fundamental branes, the D-branes and the standard solitons
are in all cases in correspondence with their potentials, the same
is not true for the non-standard solitonic branes, and indeed the
analysis of \cite{Bergshoeff:2011zk} reveals that only some
components of the representations of the solitonic fields actually
lead to supersymmetric branes. The overall result can be nicely
summarised by introducing a set of wrapping rules that give the
number of fundamental branes (F), D-branes (D) and solitons (S) in
dimension $D$ from those in dimension $D+1$
\cite{Bergshoeff:2011mh}:\,\footnote{Since there are two theories in
$D=10$\, (IIA and IIB) it is understood that the wrapping rule is
applied as follows when reducing from $D=10$ to $D=9$ dimensions:
a nine-dimensional ``undoubled'' brane can be seen
as coming from IIA and from  IIB, and consistenlty the set of
undoubled branes coming from {\it either} IIA {\it or} IIB is the same; a  nine-dimensional  ``doubled'' brane has
only one origin in terms of ten-dimensional branes, which is a IIA
or a IIB brane, and the set of doubled branes results from {\it
both} IIA {\it and} IIB, treating each resulting brane as different.
}
    \begin{eqnarray}\label{branewrapping}
 & & {\rm F} \ \ \ \left\{ \begin{array}{l}
{\rm wrapped} \ \ \ \ \ \rightarrow  \ \ {\rm doubled}\\
{\rm unwrapped} \ \ \rightarrow \ \ {\rm undoubled}
 \end{array} \right.\nonumber \\
 & & {\rm D} \ \ \ \left\{ \begin{array}{l}
{\rm wrapped} \ \ \ \ \ \rightarrow  \ \ {\rm undoubled}\\
{\rm unwrapped} \ \ \rightarrow \ \ {\rm undoubled}
 \end{array} \right.\\
 & & {\rm S} \ \ \ \left\{ \begin{array}{l}
{\rm wrapped} \ \ \ \ \ \rightarrow  \ \ {\rm undoubled}\\
{\rm unwrapped} \ \ \rightarrow \ \ {\rm doubled} \quad .
 \end{array} \right. \nonumber
\end{eqnarray}
This means that all the branes in a given dimension can be obtained
by a simple counting rule starting from the ten-dimensional ones.

The wrapping rule for fundamental branes and D-branes can be easily
understood. For fundamental branes, the doubling upon wrapping corresponds to the
fact that after compactification on a circle there is an extra
fundamental 0-brane resulting from the reduction of a pp-wave, while
for D-branes the wrapping rule simply means that the ten-dimensional
D-branes (of either IIA or IIB) generate the whole spectrum of
D-branes in any dimensions.
For {\it standard} solitons, the
doubling is precisely the dual of the one for fundamental branes,
and it corresponds to an additional contribution to the number of
solitonic $(D-4)$-branes due to a wrapped Kaluza-Klein (KK) monopole.

To realize the same dual wrapping rule for the {\it non-standard}
solitons, one needs a
class of so-called  generalised KK
monopoles with 6 worldvolume, $n $ isometry and $4-n$ transverse
directions\ $(n=0,1,2,3,4)$ \cite{Bergshoeff:2011zk}. Here $n=0$
corresponds to the NS-NS 5-brane and $n=1$ to the standard KK
monopole. Formally, one can associate to these generalised KK
monopoles the following mixed-symmetry fields:
\begin{equation}\label{d6+n,n}
\text{IIA/IIB}\,:\ \ D_{6+n,n}\,,\hskip 1truecm n=0,1,2,3,4\,.
\end{equation}
The field $D_6$ is the magnetic dual of the NS-NS 2-form $B_2$,
while the field $D_{7,1}$, which is the dual of the graviton, is
associated to the standard KK monopole. Although this dual graviton
field $D_{7,1}$  can only be introduced consistently at the
linearized level, it can still be considered as a tool to determine
all the lower-dimensional standard solitons by dimensional
reduction.
Solutions corresponding to more general mixed-symmetry
fields have been considered in e.g.~\cite{Meessen:1998qm,Englert:2003py}.
The whole set  of
supersymmetric solitons in any dimensions can be obtained from these
mixed-symmetry fields by imposing a restricted reduction rule which
states that a supersymmetric brane is only obtained when the $n$
indices on the right of the comma in $D_{6+n,n}$ are internal and
along directions that coincide with $n$ of the indices on the left
of the comma.

The only ten-dimensional supersymmetric brane which is left aside by
this analysis~\footnote{We are not taking into account the
ten-dimensional space-filling branes. These branes can only wrap.}
is the S-dual of the D7-brane of the IIB theory. The tension of this
brane scales like $(g_S)^{-3}$ in the string frame. In any dimension
below ten, one can deduce the T-duality representations of the
$\alpha=-3$ fields by simply looking at the tables in ref.~\cite{Bergshoeff:2011zk}. This leads to the remarkable result that
also for these fields the pattern of T-duality representations is
universal, see Table \ref{EfieldsanyD}.
\begin{table}[h]
\begin{center}
\begin{tabular}{|c||c|}
\hline \rule[-1mm]{0mm}{6mm} $(D-2)$-form & $E_{D-2, \dot{a}}$ \\
\hline \rule[-1mm]{0mm}{6mm} $(D-1)$-form & $E_{D-1, A\dot{a}} + E_{D-1,a}$ \\
\hline \rule[-1mm]{0mm}{6mm} $D$-form & $E_{D,AB\dot{a}} + E_{D, Aa} + 3 E_{D,\dot{a}}$
\\
 \hline
\end{tabular}
\end{center}
\caption{\sl Forms with $\alpha=-3$ in any dimension. All
representations are meant to be irreducible, and the T-duality vector indices
$AB...$ are always meant to be antisymmetrised. The $a\,, {\dot a}$ denote chiral
and anti-chiral T-duality spinor indices.
  \label{EfieldsanyD}}
\end{table}

In this letter we will analyse the structure of the WZ terms
corresponding to the fields in Table \ref{EfieldsanyD}, in order to
determine which of them correspond to supersymmetric branes. As we will see,
for the $(D-1)$- and the $D$-forms only the highest dimensional
irreducible representation corresponds to a supersymmetric brane. Moreover, we
will discover that only a subset of the components of the
representations of these fields actually corresponds to a supersymmetric brane. The
final result will lead to yet another wrapping rule:
  \begin{eqnarray}\label{newwrapping}
 & & {\rm wrapped} \ \ \ \ \rightarrow\  \ \ {\rm doubled}\,, \nonumber \\
& & {\rm unwrapped} \ \ \rightarrow \ \ {\rm doubled}\,.
 \end{eqnarray}
That is, one obtains the right counting if, going from $D+1$ to $D$
dimensions, both wrapped and unwrapped branes get doubled.

We will finally show that precisely this counting arises from
considering, together with the S-dual of the D7-brane, a specific
set of ten-dimensional objects which we generically denote as
``generalised KK monopoles''. The same result can be obtained from
the IIA point of view, in which case all the branes can be seen to
arise from compactifications of generalised KK monopoles as there is
no $\alpha=-3$ brane in IIA string theory.


\section{A new wrapping rule}
We start by reviewing how  the S-dual of the D7-brane of IIB string
theory satisfies our brane criterion, i.e.~the construction of a
gauge-invariant WZ term requires the introduction of world-volume
fields that can be associated to an eight-dimensional vector
multiplet \cite{Bergshoeff:2011zk}. Denoting with $E_8$ the $\alpha
=-3$ 8-form potential, and using the notations of
\cite{Bergshoeff:2011zk}, one obtains the field strength and gauge
transformations
  \begin{eqnarray}
  & & K_9 = d E_8 + G_3 D_6- \tfrac{1}{2} F_7 C_2\,,\nonumber \\
 & &  \delta E_8 = d \Xi_7 + G_3 \Lambda_5 - \tfrac{1}{2} F_7
 \lambda_1 \quad .
 \end{eqnarray}
Here $D_6$ is a solitonic field ($\alpha=-2$), $C_2$ is a RR field
($\alpha=-1$), $G_3$ is the curvature of $C_2$ and $F_7$ is the
curvature of $D_6$. The explicit expressions can be found in
\cite{Bergshoeff:2011zk}. Furthermore, $\Xi_7\,, \Lambda_5$ and
$\lambda_1$ are the $\alpha =-3\,,\alpha=-2$ and $\alpha=-1$ gauge
parameters. One can easily write down a corresponding WZ term, which
contains the world volume fields $c_1$ (associated to the RR field
$C_2$) and $d_5$ (associated to the solitonic field $D_6$) together
with two transverse scalars. Imposing electromagnetic duality
between $c_1$ and $d_5$ one obtains a vector plus two scalars, which
is the bosonic sector of a vector multiplet on an 8-dimensional
world volume.

We now want to repeat the same analysis in any dimension $D= 10-d$,
and determine which of the potentials in Table \ref{EfieldsanyD}
correspond to branes by analysing the world-volume field content of
the corresponding WZ term. According to our brane criterion  the
worldvolume fields have to form the bosonic sector of a vector
multiplet after imposing worldvolume electromagnetic duality and
after including the transverse scalars.

The outcome of this analysis, which we present below, will be that
the number of supersymmetric branes is
  \begin{eqnarray}
  & & (D-3)-{\rm branes} \ \ : \ \ \ 2^{d-1}\,, \nonumber \\
   & & (D-2)-{\rm branes} \ \ : \ \ \ d \times 2^{d-1}\,, \nonumber \\
  & & (D-1)-{\rm branes} \ \ : \ \ \ {d \choose 2} \times 2^{d-1} \quad .
  \label{countingrule}
\end{eqnarray}
This is summarised in Table \ref{alpha-3table} for any dimension.
\begin{table}[h]
\begin{center}
\begin{tabular}{|c||c|c|c|c|c|c|c|c|}
\hline \rule[-1mm]{0mm}{6mm} $p$-brane &IIA/IIB& 9 & 8 & 7 & 6&5&4&3\\
\hline \hline \rule[-1mm]{0mm}{6mm} 0&&&&&&&&64\\
\hline \rule[-1mm]{0mm}{6mm} 1&&&&&&&32&448\\
 \hline \rule[-1mm]{0mm}{6mm} 2&&&&&&16&192&1344\\
 \hline \rule[-1mm]{0mm}{6mm} 3&&&&&8&80&480&\\
 \hline \rule[-1mm]{0mm}{6mm} 4&&&&4&32&160&&\\
 \hline \rule[-1mm]{0mm}{6mm} 5& &&2&12&48&&&\\
 \hline \rule[-1mm]{0mm}{6mm} 6& & 1 & 4 &12&&&&\\
 \hline \rule[-1mm]{0mm}{6mm} 7& 0/1& 1 & 2 & & &&&\\
\hline
\end{tabular}
\caption{\sl By applying the wrapping rule (\ref{newwrapping})  one
obtains precisely the number of $\alpha =-3$ supersymmetric branes
predicted by the supergravity counting rule (\ref{countingrule}).
\label{alpha-3table}}
\end{center}
\end{table}
It is straightforward to realise that the numbers we get are exactly
reproduced by the wrapping rule \eqref{newwrapping}, together with
the ``initial condition'' that there is only one such brane in ten
dimensions, which is a IIB 7-brane.
\bigskip

We now proceed by deriving the counting rule  (\ref{countingrule}).
We will consider each form occurring in Table \ref{EfieldsanyD} separately, starting from the one of
lowest rank. We use the notation of \cite{Bergshoeff:2011zk}. We
thus denote with ${\cal F}_{1, A}$  the T-duality vector of
worldvolume field-strengths associated to the fundamental 1-forms
$B_{1,A}$ and to the corresponding worldvolume scalars $b_{0,A}$.
The RR fields are denoted with $C$ and their field-strengths with
$G$, while the corresponding worldvolume fields and field-strengths
are $c$ and ${\cal G}$. All these objects are in spinor
representations of the T-duality group $\text{SO}(d,d)$ of
alternating chirality. The solitonic fields that we consider are the
fields $D_{D-4+i, A_1 ...A_i}$ for $i=0,...,4$, and we denote their
field strengths with $H$. We associate to these fields the
worldvolume fields $d_{D-5+i, A_1 ...A_i}$, with field strength
${\cal H}_{D-4+i, A_1 ...A_i}$. Finally, $\Gamma_A$ denotes the
Gamma matrices of the T-duality group. We refer to the Appendix of
\cite{Bergshoeff:2011zk} for all the properties of these Gamma
matrices that will be relevant in the analysis below.

\subsection{$(D-2)$-forms}
The $\alpha=-3$ $(D-2)$-forms always belong to the irreducible spinor
representation denoted by the lower index $\dot{a}$, which is the
same chirality as the RR 2-forms $C_{2,\dot{a}}$. We want to
determine whether one can write down a WZ term associated to this
field that contains the right number of world volume degrees of
freedom to form the bosonic sector of a half-supersymmetric vector
multiplet. Together with the two transverse scalars resulting from a
$D-2$ dimensional world volume in $D$ dimensions, one needs in
addition $d$ scalars and one vector. This makes a total of $d+2$,
that is $10 - (D-2)$ scalars as appropriate to a $D-2$ dimensional
vector multiplet.

We schematically write down the WZ term without computing the actual
coefficients. This will turn out in all cases to be enough to
determine the supersymmetric branes.  The WZ term is
  \begin{equation}
  E_{D-2} + \sum_{i=0}^1 a_i D_{D-4+i, \{ A_i \} } \Gamma^{\{A_i \}
  } {\cal G}_{2-i} + \sum_{i=0}^1 b_i \Gamma^{\{ A_i \}
  } C_{2-i} {\cal H}_{D-4+i , \{ A_i \} }
\quad ,
\end{equation}
where in general $\{A_i \}$ denotes $i$ antisymmetric
$\text{SO}(d,d)$ vector indices, while all the T-duality spinor
indices are understood. Moreover, ${\cal G}$ are the field strengths
of the $\alpha=-1$ world volume fields $c_{2n,a}$ and $c_{2n+1,
\dot{a}}$ and ${\cal H}$ are the field strengths of the $\alpha=-2$
world volume fields $d_{D-5}$ and $d_{D-4,A}$. We now count the degrees of freedom,
assuming that all the coefficients $a_i$  and $b_i$ are
non-vanishing (this will be the assumption that we will make
throughout this section). The terms proportional to $a_0$ and $b_0$
propagate the fields $c_{1, \dot{a}}$ (the index ${\dot{a}}$ is
fixed) and $d_{D-5}$, which corresponds to a vector and its dual.
The terms proportional to $a_1 $ and $b_1$ propagate the scalars $c_{0,a}$ and
their duals $d_{D-4,A}$. To do the counting, one has to perform a light-cone
Gamma matrix analysis similar to the one of
\cite{Bergshoeff:2011zk}.
Following \cite{Bergshoeff:2011zk} we use a light-cone basis
$\Gamma_{n \pm}$ for the Gamma matrics. Given that the index $\dot{a}$ is fixed, one can
show that for each $n$ only one non-vanishing Gamma matrix appears
in the WZ term. This means that in the term proportional to $b_1$
one has to count only half of the $2d$ indices, which makes $d$
fields $d_{D-4}$. The same applies for the term proportional to
$a_1$: the non-vanishing Gamma matrices project the $2^{d-1}$
components of the field $c_{0,a}$ to $d$ independent components.
Imposing electromagnetic duality between the $c_{0,a}$ and the $d_{D-4,A}$ fields,
one is left with $d$ scalars. The conclusion is that we expect all
the components of the field $E_{D-2,\dot{a}}$ to be associated to
supersymmetric branes.

\subsection{$(D-1)$-forms}

We now consider the $(D-1)$-forms of Table \ref{EfieldsanyD}. It is
immediately apparent that the field $E_{D-1,a}$ can never satisfy
our criteria since its corresponding WZ term contains far too many
worldvolume fields (and in particular it contains the 2-form $c_2$
which cannot be included in a vector multiplet in general). We are
thus led to consider only the field $E_{D-1, A \dot{a}}$ in the
irreducible ``gravitino'' representation of T-duality. The most
general WZ term for this field is
  \begin{eqnarray}
  & & E_{D-1,A} + \sum_{i=0}^1 c_i D_{D-3+i,A \{ A_i \} } \Gamma^{\{A_i \}
  } {\cal G}_{2-i}+ \sum_{i=0}^1 \tilde{c}_i D_{D-3+i,\{ A_{i+1} \} } \Gamma_{A}{}^{ \{A_{i+1} \}
  } {\cal G}_{2-i} \nonumber \\
  & & + \sum_{i=0}^1 d_i \Gamma^{\{ A_i \}
  } C_{2-i} {\cal H}_{D-3+i , A\{ A_i \} } + \sum_{i=0}^1 \tilde{d}_i \Gamma_A{}^{\{ A_{i+1} \}
  } C_{2-i} {\cal H}_{D-3+i , \{ A_{i+1} \} }\nonumber\\
  &&
  + E_{D-2} {\cal F}_{1,A} - \frac{1}{2d-1} \Gamma_{AB} E_{D-2}
  {\cal F}_1^B \label{ED-1WZterm}
\quad ,
\end{eqnarray}
where the coefficients $c_i$ and $\tilde{c}_i$, as well as $d_i$ and
$\tilde{d}_i$, are related so that the resulting expression is
Gamma-traceless, and the first term in the last line has been
normalised to 1, as one can always do up to field redefinitions.

We now want to count the worldvolume degrees of freedom.  We first
count the vectors, that correspond to the terms proportional to
$c_0$, $\tilde{c}_0$, $d_0$ and $\tilde{d}_0$. The terms $c_0$ and
$d_0$  propagate a single vector $c_{1,\dot{a}}$ and a single
$(D-4)$-form $d_{D-4, A}$, which is dual to a vector (the indices $\dot{a}$
and $A$ are fixed).
We are going to show below  that for a given set of
lightlike components inside the gravitino representation these two
terms are automatically Gamma-traceless, so that the terms
$\tilde{c}_0$ and $\tilde{d}_0$ are not needed. More precisely,
 both terms $c_0$ and $\tilde{c}_0$ in the
Minkowskian base contribute to give the single term in the lightcone
base, and similarly for the other two terms. The absence of
these terms guarantees that only one worldvolume vector propagates.
It will turn out that these components are exactly those that
propagate the right amount of scalars.

To prove the statement above it is convenient to use  lightcone coordinates. For each lightlike direction
$n\pm$, the corresponding Gamma matrix $\Gamma_{n \pm, \dot{a}}{}^a$
is vanishing for half of the values of $\dot{a}$ and non-vanishing
for the other half. We take the components of $E_{D-1, A\dot{a}}$ to be
along the directions for which the corresponding Gamma matrix has
only vanishing entries. This forms a $d \times 2^{d-1}$ dimensional
orbit within the gravitino representation. If for instance we take
the component $E_{D-1, n+ \dot{a}} $ such that the matrix
$\Gamma_{n+, \dot{a}}{}^a$ vanishes, then the matrix $\Gamma_{n-, a
}{}^{\dot{a}}$ vanishes too, which implies that the term $c_0$ and
the term $d_0$ are automatically Gamma-traceless along these
components.
This completes the proof of the statement.

We now count the scalars. We first consider the term $c_1$. If the
index $A$ of $D_{D-2,AB}$ is say $1+$, the index $B$ can be $1-$ or
any or the other $n\pm$ indices, with $n\neq 1$. But if $B = 1-$,
then the Gamma matrix in the $c_1$ term is $\Gamma_{1+,
\dot{a}}{}^a$ which is vanishing for the $\dot{a}$ we are
considering. For all the other possibilities, for each $n$ there is
always one and only one of the two possibilities $+$ or $-$ for
which the corresponding Gamma matrix is non-vanishing. This makes in
total $d-1$ possibilities, and for each possibility one picks a
scalar field $c_{0, a}$. One thus selects $d-1$ out of the $2^{d-1}$
scalars. Analogously, for the $d_1$ term one selects the $(D-3)$-forms
$d_{D-3, 1+ B}$ such that $B$ is not $1-$ and is only one
possibility out of $n \pm$ for each $n\neq 1$. These are $d-1$
$(D-3)$-forms which are dual to the scalars. Finally, there are two
additional scalars. One is the transverse scalar corresponding to a
$(D-1)$-dimensional world volume in $D$ dimensions. The other is  $b_{0,A}$
for fixed index $A$. The previous argument shows again that
in lightcone notation and for the lightcone components we are
considering the last term in \eqref{ED-1WZterm} should not be
written. We thus have a total of $d+1=11-D$ worldvolume scalars,
which is the correct amount for a $(D-1)$-dimensional worldvolume. To
summarise, the number of supersymmetric branes is
  \begin{equation}
  d \times 2^{d-1} \quad .
\end{equation}

\subsection{$D$-forms}
We now consider the $D$-forms, corresponding to the last line in
Table \ref{EfieldsanyD}. Again, as in the previous case, it is
straightforward to see that only the highest dimensional irreducible
tensor-spinor representation can lead to the right worldvolume
fields. We thus consider the WZ term
    \begin{eqnarray}
  & & E_{D,AB} + \sum_{i=0}^1 e_i D_{D-2+i,AB \{ A_i \} } \Gamma^{\{A_i \}
  } {\cal G}_{2-i}\nonumber\\
  &&+ \sum_{i=0}^1 f_i \Gamma^{\{ A_i \}
  } C_{2-i} {\cal H}_{D-2+i , AB\{ A_i \} }
  + E_{D-1, [A} {\cal F}_{1,B]}  \label{EDWZterm}
\quad ,
\end{eqnarray}
where it is understood that each term is projected on its
Gamma-traceless part.

We now want to determine the components that give rise to a
worldvolume vector multiplet. We consider the indices $AB$ to be of
the form $n\pm m\pm$ with $n \neq m$. We take for simplicity the
direction $1+2+$. We consider the Gamma matrices
$\Gamma_{1+,\dot{a}}{}^a$ and $\Gamma_{2+,\dot{a}}{}^a$, and we take
the directions $\dot{a}$ such that both $\Gamma_{1+}$ and
$\Gamma_{2+}$ vanish. These directions are one fourth of the
original spinor components, that is $2^{d-3}$ directions.
We wish to
show that for each of these directions the corresponding WZ term
propagates the right degrees of freedom. This gives a total number
of branes equal to
  \begin{equation}
{d \choose 2} \times 2^{d-1} \quad .
\end{equation}

We first consider the vector. This arises from the terms $e_0$ and
$f_0$. Given that the index $\dot{a}$ and the indices $AB$ are
fixed, this clearly propagates one vector and its dual. What remains
to be seen is that for the components we have selected this is
automatically Gamma-traceless. This is automatic, because the Gamma
trace corresponds to contracting with $\Gamma_{1-,a}{}^{\dot{a}}$ or
$\Gamma_{2-, a}{}^{\dot{a}}$, which is identically zero for the
values of $\dot{a}$ that we have selected. What remains to be
considered are the scalars. This corresponds to the $e_1$ and the
$f_1$ terms. In both terms, the index $C$ in $ABC$ can be $1-$, $2-$
or any $m\pm$ with $m\neq 1,2$. But in the first two cases, the
corresponding Gamma matrix in the WZ term vanishes, so the only
possibility is the third, and actually for each $m$ there is only
one of the two possibilities $m+$ or $m-$ that gives a non-vanishing
result. This selects $d-2$ possibilities. In the $e_1$ term, the
$d-2$ Gamma matrices project on $d-2$ independent combinations of
scalars out of the $2^{d-1}$ scalars $c_{0,a}$, while in the $f_1$
term this simply selects $d-2$ fields $d_{D-2, 1+2+ m}$ which are
dual to the scalars. To these $d-2$ scalars we have to add the two
scalars $b_{0,A}$ and $b_{0,B}$. This gives $d=10-D$ scalars, which
leads to the right number of degrees of freedom for a
$D$-dimensional worldvolume.
 It is easy to show that all the
terms involving the scalars are automatically Gamma traceless for
the components we have selected.
\bigskip

This concludes our proof of the counting rule (\ref{countingrule})
which is in line with the new wrapping rule (\ref{newwrapping}).

\section{Generalised KK monopoles}
As mentioned in the introduction the realization of the soliton
wrapping rule \eqref{branewrapping} requires the introduction of a
set of generalized KK monopoles together with the solitonic 5-brane
and the standard KK monopole \cite{Bergshoeff:2011mh}. One can
associate the mixed-symmetry fields given in (\ref{d6+n,n}) to these
generalized monopoles. Applying a restricted reduction rule to these
mixed-symmetry fields yields precisely the same number of solitions
that follows form our supergravity analysis.

We now want to perform a similar analysis for the $\alpha=-3$
branes. In particular, we wish to determine which extra objects,
which we will generically  denote by ``generalized KK monopoles'',
are needed to realize the new wrapping rule (\ref{newwrapping}). We
find that all the branes in Table \ref{alpha-3table}, satisfying the
wrapping rule \eqref{newwrapping}, can be obtained from the
following set of ten-dimensional mixed-symmetry fields
  \begin{eqnarray} & & {\rm IIA} \ \quad E_{8+n, 2m+1 , n} \quad
  \quad n=0,1,2 \qquad 2m+1 \geq n\,, \label{E8+n2m+1nIIA} \\
  & & {\rm IIB} \ \quad E_{8+n, 2m , n} \quad
  \quad \hskip .45truecm n=0,1,2 \qquad 2m \geq n \quad , \label{E8+n2mnIIB}
  \end{eqnarray}
provided that one uses a similar restricted compactification rule as
in the case of the solitons. Explicitly, we have the IIA fields
  \begin{eqnarray} & & E_{9,1,1} \quad E_{8,1} \quad E_{10,3,2}
  \quad E_{9,3,1} \quad E_{8,3} \quad E_{10,5,2} \quad E_{9,5,1}
  \quad E_{8,5} \quad E_{10,7,2} \nonumber \\&& E_{9,7,1} \quad E_{8,7}
  \end{eqnarray}
and the IIB fields
  \begin{eqnarray} & & E_{8} \quad E_{10,2,2}
  \quad E_{9,2,1} \quad E_{8,2} \quad E_{10,4,2} \quad E_{9,4,1}
  \quad E_{8,4} \quad E_{10,6,2} \nonumber \\&& E_{9,6,1} \quad E_{8,6} \quad .
  \end{eqnarray}

As an example we  show how the counting works in seven dimensions. We have,
from IIA,
   \begin{eqnarray}
  & & E_{9,1,1} \rightarrow E_{6ijk,i,i} \ (3) \quad E_{7ij,i,i} \
  (6)\,,
  \nonumber\\
  & & E_{8,1} \rightarrow E_{5ijk,i} \ (3) \quad E_{6ij,i} \ (6) \quad E_{7i,i} \ (3)\,, \nonumber
  \\
  & & E_{10,3,2} \rightarrow E_{7ijk,ijk,ij} \ (3)\,,
  \nonumber \\
  & & E_{9,3,1} \rightarrow E_{6ijk,ijk,i} \ (3)\,, \nonumber \\
  & & E_{8,3} \rightarrow E_{5ijk,ijk} \ (1) \quad ,
  \end{eqnarray}
where we have used the restricted reduction rule that in
$E_{m,n,p}$, with $m \ge n \ge p$,  all $p$ indices must be internal
and that these internal indices must also occur among the $m$ and
$n$ indices.
Furthermore,  the remaining $n-p$ indices among the $n$ indices are also taken to be internal,  and these must also occur among the $m$ indices.
For the $E_{m,n}$ fields we use the same restricted
reduction rule as for the solitons, see the introduction. Applying these restricted reduction rules  gives
four 4-branes, twelve 5-branes and twelve 6-branes, which is the correct
result, cp.~to Table 2. One can easily show that the IIB
compactification gives the same result. Similarly, one can show that
all the other dimensions work in the same way.

\section{Conclusions}

In this letter we showed, by completing our earlier work, that
branes whose tension scales as $T \sim (g_S)^\alpha$ for $\alpha
=0,-1,-2,-3$ satisfy the following wrapping rule
\begin{equation}\label{simple}
 \begin{array}{l}
{\rm wrapped} \ \ \ \ \ \rightarrow\   \ {\rm doubled}\,,  {\rm undoubled}\,, {\rm undoubled}\,, {\rm doubled}\,,\\
{\rm unwrapped} \ \ \rightarrow \ \ {\rm undoubled}\,, {\rm undoubled}\,, {\rm doubled}\,,  {\rm doubled}\,,
 \end{array}
 \end{equation}
where the four terms at the right of the arrow correspond to
$\alpha=0,-1,-2$ and $-3$, respectively. For $\alpha=0$ the doubling
of branes is due to the reduction of pp-waves. Dirichlet branes,
with $\alpha=-1$, have no doubling and are complete by themselves.
For standard solitonic branes, with $\alpha=-2$, the doubling is due
to the presence of  the standard KK monopole. In our previous paper
\cite{Bergshoeff:2011mh} we suggested that the doubling in the case
of non-standard solitons is due to the presence of so-called
generalized KK monopoles. Similarly, in the present paper we
introduced a new wrapping rule for $\alpha=-3$ and suggested that
the doubling is due to the presence of new objects which we
generically called generalized KK monopoles.

At present it is not clear what the precise status of the generalized KK
monopoles is. We are able to associate a  set of mixed-symmetry
fields to them with a restricted reduction rule such that all branes
suggested  by supergravity are generated upon reduction. The
explicit solution for some of the suggested generalized KK monopoles have been given in the literature, see
e.g.~\cite{Meessen:1998qm,Englert:2003py}. What is not yet clear is whether a finite energy solution
can be obtained, possibly by taking superpositions of such generalized KK monopoles.
In the
introduction we stated that supergravity is incomplete in the sense
that the maximal supergravity theories in different dimensions are
not related to each other by toroidal reduction. In some sense the
new structure we introduced, generalized KK monopoles or
mixed-symmetry fields, takes this incomplete nature of supergravity
away. Whether this is merely a book keeping trick or a true
physical meaning can be given to the generalized KK monopoles
remains to be explored. The role of the very extended Kac-Moody
algebra $E_{11}$ \cite{West:2001as} in this is intriguing. Not only
does $E_{11}$ predict the number of physical and un-physical
potentials of maximal supergravity, it also contains as a sub-sector
the mixed-symmetry fields \eqref{d6+n,n}, \eqref{E8+n2m+1nIIA} and
\eqref{E8+n2mnIIB} associated to the generalized KK monopoles.

Ten-dimensional string theory does not contain branes with $\alpha
< -4$. The IIB theory contains a space-filing brane with
$\alpha=-4$, the S-dual of the D9-brane, but space-filling branes
can only wrap and therefore no non-trivial wrapping rule can be
associated with them. Indeed, for $\alpha=-4$ we do not find a
visible pattern like for the higher values of $\alpha$.
Interestingly, lower-dimensional maximal supergravity suggests the
existence of non-space-filling branes with $\alpha=-4$. For
instance, in $D\le 6$ dimensions there are domain walls with
$\alpha=-4$ and in $D=3, 4$ dimensions there are branes of
co-dimension 2 with $\alpha=-4$. Clearly, such branes do not follow
from the reduction of the ten-dimensional IIB space-filling brane
and must be the result of reducing a generalized KK monopole with
$\alpha=-4$. Similarly, in $D\le 6$ dimensions maximal supergravity
suggests branes with $\alpha\le -5$ and such branes too must be the
result of generalized KK monopoles with $\alpha\le -5$.

Summarizing, we find that all branes of IIA and IIB string theory,
excluding the space-filling branes which should be treated
separately, satisfy the  wrapping rule \eqref{simple}. The deeper
meaning of why branes should satisfy such a simple wrapping rule is
unclear to us. It would be interesting to see whether some
geometrical  interpretation could be given of this  rule. In this
respect it would be interesting to investigate the doubled wrapping
rule we find for the S-dual of the D7-brane and to see whether this
could be understood from an F-theory \cite{Vafa:1996xn} point of
view.

\section*{Acknowledgements}
We would like to thank the organisers of the ``Istanbul 2011:
strings, branes and supergravity'' conference for providing a very
stimulating environment where part of this work was carried out.
F.R. would like to thank the University of Groningen for hospitality
at the early stages of this work. E.B. would like to thank King's
College for its hospitality.






\end{document}